\documentclass{article}
\usepackage{spconf,amsmath,graphicx}
\usepackage{amsfonts,amssymb}
\usepackage{float}
\usepackage{array}
\usepackage{booktabs}
\usepackage{epstopdf}
\usepackage{color}
\usepackage{hyperref}
\usepackage{stfloats}

\title{DSPGAN: a GAN-based universal vocoder for high-fidelity TTS by time-frequency domain supervision from DSP}
\name{Kun Song$^1$, Yongmao Zhang$^1$, Yi Lei$^1$, Jian Cong$^1$, Hanzhao Li$^1$, Lei Xie$^{1*}$, Gang He$^2$, Jinfeng Bai$^2$}
\vspace{-5pt}
\address{
  $^1$Audio, Speech and Language Processing Group (ASLP@NPU)\\School of Computer Science,
  Northwestern Polytechnical University, Xi’an, China \\
  $^2$TAL Education Group, Beijing, China}
\vspace{-5pt}
\vspace{-10pt}
\begin{document}
\vspace{-5pt}
\ninept
\maketitle
\begin{abstract}
\vspace{-5pt}
Recent development of neural vocoders based on the generative adversarial neural network (GAN) has shown obvious advantages of generating raw waveform conditioned on mel-spectrogram with fast inference speed and lightweight networks. Whereas, it is still challenging to train a universal neural vocoder that can synthesize high-fidelity speech from various scenarios with unseen speakers, languages, and speaking styles. In this paper, we propose DSPGAN, a GAN-based universal vocoder for high-fidelity speech synthesis by applying the time-frequency domain supervision from digital signal processing (DSP). To eliminate the mismatch problem caused by the ground-truth spectrograms in the training phase and the predicted spectrograms in the inference phase, we leverage the mel-spectrogram extracted from the waveform generated by a DSP module, rather than the predicted mel-spectrogram from the Text-to-Speech (TTS) acoustic model, as the time-frequency domain supervision to the GAN-based vocoder. We also utilize sine excitation as the time-domain supervision to improve the harmonic modeling and eliminate various artifacts of the GAN-based vocoder. Experiments show that DSPGAN significantly outperforms the compared approaches and it can generate high-fidelity speech for various TTS models trained using diverse data. \footnote{Audio samples are available at \url{https://kunsung.github.io/DSPGAN/}}
\end{abstract}

\vspace{-5pt}
\vspace{-3pt}
\renewcommand{\thefootnote}{\fnsymbol{footnote}}
\footnotetext{Corresponding author. This work was supported by National Key R\&D Program of China, under Grant No. 2020AAA0104500.}

\vspace{-5pt}
\section{Introduction}
\label{sec:intro}
\vspace{-7pt}
Although end-to-end systems have gradually come to the stage~\cite{DBLP:conf/iclr/DonahueDBES21, DBLP:conf/icml/KimKS21}, the current mainstream Text-to-Speech (TTS) systems based on neural networks are still a two-stage architecture composed of an acoustic model and a neural vocoder due to its flexibility in deployment. 
The acoustic model produces intermediate representations such as mel-spectrogram from text, and the neural vocoder uses the predicted intermediate representations as input to generate speech waveform. To ensure good quality, both the acoustic model and vocoder are trained by a great number of high-quality data of target speakers. In diverse scenarios of real applications, it's critical to build a universal vocoder for flexibly generating high-quality speech without any fine-tuning from arbitrary acoustic models of various speakers, speaking styles, and even languages.
But building a universal neural vocoder is not a trivial task, due to several challenges. Specifically, the inference of unseen data may lead to an inferior performance compared with the seen data during training, which makes the robustness of the unseen data critical to the universal vocoder. Besides, the distribution mismatch between the imperfectly predicted and the ground truth intermediate representations also leads to audio quality degradation with artifacts.

This paper aims to design a robust universal neural vocoder without any fine-tuning process for producing high-quality speech with mel-spectrogram predicted from various acoustic models, unseen speakers, speaking styles, and languages.
To achieve our goal, we focus on the GAN-based vocoder~\cite{DBLP:conf/nips/KumarKBGTSBBC19, DBLP:conf/icassp/YamamotoSK20, DBLP:conf/nips/KongKB20} due to its good performance with fast inference speed and lightweight networks. GAN-based vocoders have become a mainstream non-autoregressive (non-AR) neural vocoder because they can synthesize high-fidelity waveform with faster training and inference speed than flow-based~\cite{DBLP:conf/icassp/PrengerVC19, DBLP:conf/icml/PingPZ020} and diffusion-based~\cite{DBLP:conf/iclr/KongPHZC21} vocoders. However, to build a universal vocoder, GAN-based vocoders still have the following problems. 
First, the GAN-based vocoder may induce prediction errors of periodic components of speech due to its unstable parallel-generating nature. These errors result in artifacts such as pitch jitters and discontinuous harmonics, especially in prolonged soundings. Second, the acoustic model may generate an over-smoothed spectrogram~\cite{DBLP:conf/acl/0006TQZL22} that has unseen during the training phase of the vocoder, leading to metallic or noisy sounding with voiced/unvoiced errors, which usually exists in highly expressive speech. Furthermore, the poor generalization of GAN-based vocoder may degrade the speech quality for unseen speakers and languages.
\begin{figure*}[ht]

	\centering
	\includegraphics[width=18.2cm,height=4.9cm]{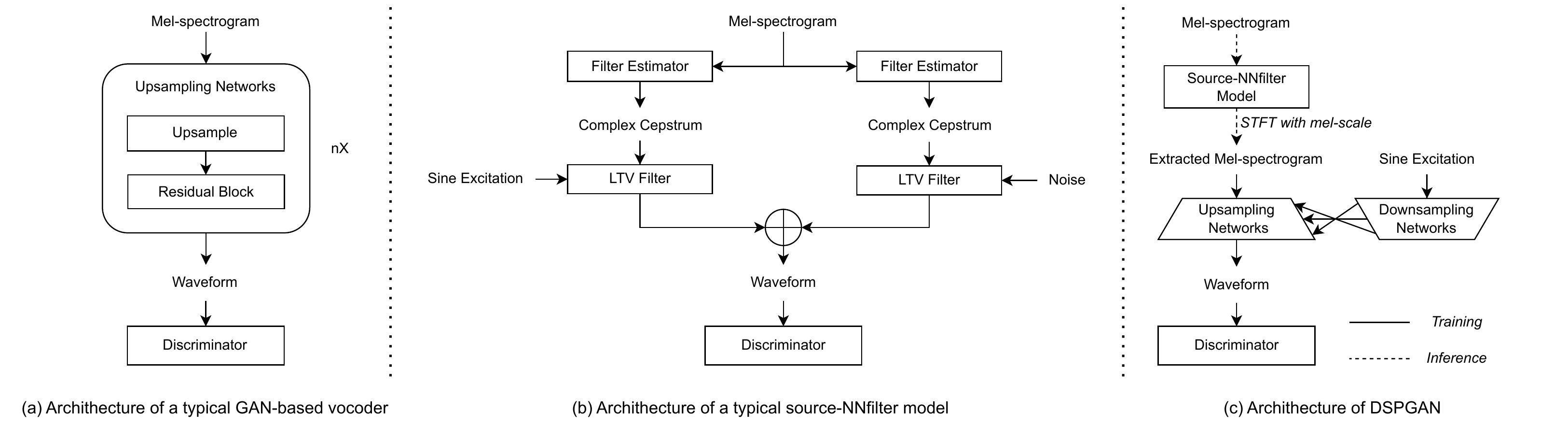}
	\vspace{-20pt}
	\caption{
		Architecture of our proposed approach.
	}
	\vspace{-15pt}
	\label{model_dspgan}
\end{figure*}
\vspace{-2pt}

A GAN-based vocoder can be regarded as a generative model that uses the intermediate representation such as mel-spectrogram as supervision. Some works try to strengthen supervision to alleviate the above problems. A straightforward trick is to use the generated mel-spectrogram from acoustic model to fine-tune the neutral vocoder biasing to the specific targets. But this fine-tuning process is time-consuming and apparently against the goal of a universal vocoder. Some studies use pitch as additional time-domain supervision in the upsampling network to solve the problems of pitch jitter and discontinuous harmonics in singing voice synthesis (SVS)~\cite{DBLP:conf/icassp/GuoZML22, DBLP:conf/interspeech/XuZG22}.
But unfortunately, with the supervision from  pitch, it is likely to generate excessive periodic components at the high frequency of speech, leading to mechanical sounding, due to the aforementioned mismatch between the ground truth and predicted spectrogram by the acoustic model. 
Moreover, apparently, supervision from pitch can only guide the generation of periodic components of speech but not the aperiodic components. As a result, with an over-smoothed spectrogram predicted from an acoustic model, metallic or noisy sounding will occur. 


Different from the data-driven rationale of neural vocoders, traditional digital signal processing (DSP) vocoders~\cite{DBLP:journals/speech/KawaharaMC99, DBLP:journals/ieicet/MoriseYO16} are designed mathematically based on the source-filter model of human vocal system. Since the periodic and aperiodic components of speech are modeled according to human vocal characteristics, DSP vocoders have strong robustness to unseen spectrograms and reasonable speech parametric controllability. Thus it is rare to induce periodic errors which result in pitch jitter and discontinuous harmonics or aperiodic errors which lead to metallic or noisy sounding. But as an over-simplified model of the human vocal system, DSP vocoders are clearly inferior to data-driven-based neural vocoders in speech quality. The synthetic speech sounds generally mechanical, caused by the strict supervision from the sine excitation. There has been a recent trend to leverage the advances from both kinds of vocoders. For example, LPCNet~\cite{DBLP:conf/icassp/ValinS19} uses linear predictive coding to simplify AR neural vocoder's architecture; NSF~\cite{DBLP:conf/icassp/WangTY19} and DDSP~\cite{DBLP:conf/iclr/EngelHGR20} introduce neural network filters to the source-filter model, and on this basis, NHV~\cite{DBLP:conf/interspeech/LiuCY20} specifically adopts GAN to optimize the waveform generation. 


Following this trend, in this paper, we propose \textit{DSPGAN} as a universal vocoder that adopts a DSP module to guide the learning of a GAN-based vocoder. Specifically, besides using sine excitation in the time domain as supervision, we also adopt extracted mel-spectrogram from the waveform synthesized by a DSP module to supervise the learning of a GAN-based vocoder in the time-frequency domain given the robustness of a DSP vocoder on the generation of periodic and aperiodic components. With better supervision, the fine details of speech are modeled from the data-driven neural vocoder without artifacts, such as metallic and noisy sounding. Experiments show the robustness of DSPGAN on speech generation from a variety of mel-spectrogram, including various acoustic models, unseen speakers, multiple speaking styles as well as multiple languages.

\vspace{-8pt}
\section{Proposed Approach}
\label{sec:pagestyle}
\vspace{-7pt}
In this section, we propose to apply time-frequency domain supervision from a DSP vocoder to a typical GAN-based neural vocoder for the purpose of a universal vocoder. To obtain more stable and higher quality supervision from DSP, instead of using a conventional source-filter model, we introduce a source neural network filter (source-NNFilter) model which adopts the learnable neural network filter; on this basis, we further make improvements to the source-NNFilter model.

\vspace{-12pt}
\subsection{A typical GAN-based vocoder}
\vspace{-5pt}
As shown in Figure~\ref{model_dspgan}(a), a typical GAN-based vocoder generates waveform by a few upsampling network layers which contain transposed convolution and a stack of residual blocks with dilated convolutions. Meanwhile, a GAN-based vocoder usually uses multiple discriminators for adversarial training to learn different frequency-domain features of speech. In addition, some auxiliary losses are usually adopted to stabilize training, such as feature matching loss and reconstruction loss. Therefore, the training objectives of its generator and discriminator are
\vspace{-5pt}
\begin{equation}
\mathcal{L}_{G}=\mathbb{E}_{\hat{x}}\left[\left(D(\hat{x})-1\right)^{2}\right] + \mathcal{L}_{recon}(G) + \mathcal{L}_{fm}(G)
\end{equation}
\vspace{-5pt}
and
\begin{equation}
\mathcal{L}_{D}=\mathbb{E}_{x}\left[\left(D(x)-1\right)^{2}\right] + \mathbb{E}_{\hat{x}}\left[D(\hat{x})^{2}\right]
\end{equation}
respectively, where $x$ and $\hat{x}$ represent ground-truth and generated samples, and $\mathcal{L}_{recon}$ and $\mathcal{L}_{fm}$ denote reconstruction loss and feature matching loss respectively.


\vspace{-8pt}
\subsection{Improved source-NNFilter model}
\vspace{-4pt}
A typical source-filter model uses sine excitation and noise as source signals respectively for the generation of the periodic and aperiodic components of speech, and linear time-varying (LTV) filters are employed to filter the source signals. Specifically in this paper, we adopt NHV~\cite{DBLP:conf/interspeech/LiuCY20} as our DSP vocoder, using neural networks to control LTV filters, which results in a source-NNFilter model with low computational complexity. As is shown in Figure~\ref{model_dspgan}(b), the source-NNFilter model uses NN filter estimators to predict complex cepstrum from the input mel-spectrogram as the impulse response. By using complex cepstrum to filter the source signals, the source-NNFilter model can dynamically control the amplitudes of different harmonic components, so it can generate speech with higher quality than the conventional source-filter model. And for sine excitation in the source-NNFilter model, it can convert from pitch via additive synthesis. 
We can firstly get the sample-level f0 $f_{0}[n]$ by linearly interpolating the frame-level f0. The corresponding sine excitation $p_{K}[n]$ generation process from $f_{0}[n]$ is
\vspace{-4pt}
\begin{equation}
\label{pn}
p_{K}[n]=\left\{\begin{array}{cc}
\sum_{k=1}^{min(K, f{s}/2f_{0}[n])} \sin \left(\phi_k[n]\right) & f_0[n]>0 \\
0 & f_0[n]=0
\end{array}\right.
\end{equation}
\vspace{-1pt}
where $K$ is the number of sine functions set in the additive synthesis. In this paper we set $K$ to 200, which means the number of harmonics is 200. Here $\phi_k[n]$ is the phase of the $k$-th harmonic at timestep $n$, calculated by
\vspace{-5pt}
\begin{equation}
\phi_k[n]=\phi_k[n-1]+2 \pi k \frac{f_0[n]}{f s}.
\end{equation}
\vspace{-1pt}
Since the pitch is required for sine excitation, to fit all acoustic models, we use a pitch predictor model independently. The pitch predictor, optimized by L2 loss, predicts the pitch from the raw mel-spectrogram over several Conv1D layers.

\begin{figure}[t]
\setlength{\abovecaptionskip}{0.1cm}
	\centering
	\includegraphics[width=0.9\linewidth]{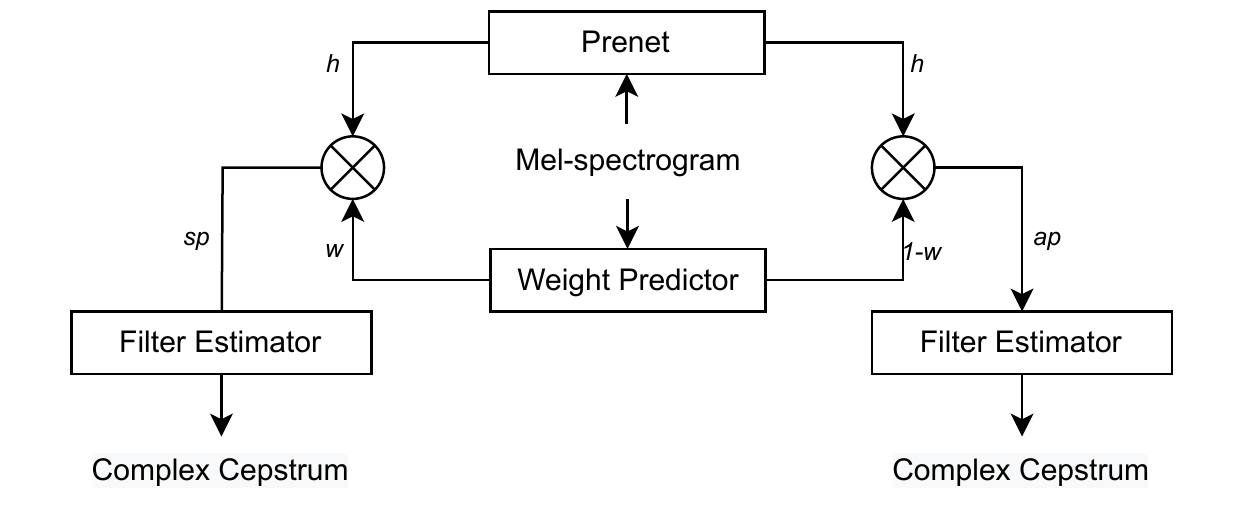}
	\vspace{-10pt}
	\caption{
		Complex cepstrum generation with weight predictor.
	}
	\vspace{-15pt}
	\label{weight}
\end{figure}

As periodic and aperiodic components are generated separately, the current source-NNFilter model may induce aberrant points in the synthesized aperiodic components, especially in the stochastic part of voiced sound related to the periodic components. Therefore, it is necessary to establish dependence between the aperiodic and periodic components to enhance the controllability of the aperiodic components and stabilize the model. As shown in Figure~\ref{weight}, we introduce a weight predictor to achieve this goal. The model predicts the hidden feature $h$ and weight $w$ from mel-spectrogram through the prenet and the weight predictor, respectively, while taking $w$ and 1-$w$ as the weights of the periodic and aperiodic filter estimator's input $sp$ and $ap$. In this way, the sum of periodic and aperiodic components matches the input mel-spectrogram in the frequency domain. The calculation process of $sp$ and $ap$ is  
\begin{equation}
\vspace{-2pt}
\left\{\begin{array}{l}
s p_{c, t}=w_{c, t} * h_{c, t} \\
a p_{c, t}=\left(1-w_{c, t}\right) * h_{c, t}
\end{array}\right.
\vspace{-2pt}
\end{equation}
where $t$ and $c$ denote the frame and channel for each feature respectively.
\vspace{-8pt}


\vspace{-5pt}
\subsection{Supervision to GAN-based Vocoder}
\vspace{-5pt}
Based on the improved source-NNFilter model, we specifically use its robust output as supervision to guild the learning of the GAN-based vocoder.

\vspace{-12pt}
\subsubsection{Supervision in time-frequency domain}
\vspace{-5pt}
For a universal vocoder, we expect it to have strong robustness to mel-spectrogram generated from the acoustic model of unseen speakers, speaking styles, and languages. But an obvious mismatch occurs between inference and training of a neural vocoder, resulting in various types of artifacts and degraded speech quality. Therefore, it is necessary to eliminate such mel-spectrogram mismatch. As shown in Figure~\ref{model_dspgan}(c), we simply adopt the mel-spectrogram of synthesized speech from a pre-trained source-NNFilter model to replace the original mel-spectrogram during training and inference. Due to the source-NNFilter model's high robustness to the generated mel-spectrogram from the acoustic model, there is no mismatch in training and inference, which can avoid various artifacts in waveform generation. 

\vspace{-12pt}
\subsubsection{Supervision in time domain}
\vspace{-5pt}
Sine excitation applied to a GAN-based vocoder to avoid periodic errors has already been proven effective in SVS~\cite{DBLP:conf/icassp/GuoZML22, DBLP:conf/interspeech/XuZG22}. But mechanical sounding similar to that which results from a source-filter model may occur due to the mel-spectrogram mismatch. Since we have solved the mismatch with time-frequency domain supervision introduced above, we can directly use sine excitation for time domain supervision. Specifically, we use a UNet-like strategy as shown in Figure~\ref{model_dspgan}(c). Sine excitation is downsampled at different scales by downsampling networks and concatenated to the hidden features of each upsampling network layer. In this way, the learning of the upsampling networks (for modeling harmonics) is supervised by sine excitation. Here we only use the pitch as sine excitation in the upsampling networks, equivalent to
 $p_{1}[n]$ in Eq.~(\ref{pn}),  
since pitch is sufficient for improving the accuracy of harmonic modeling in a GAN-based vocoder.


\vspace{-8pt}
\section{Experiments}
\label{sec:exp}
\vspace{-5pt}
\subsection{Data Configuration}
\vspace{-5pt}
All audio samples are resampled to 24KHz and 80-band log-mel-spectrogram is extracted with a 1024-point FFT, 256 sample frameshift, and 1024 sample frame length.

\vspace{-10pt}
\subsubsection{Training set}
\vspace{-4pt}
We use Mandarin and English speech as the training set. The Mandarin recordings are from an internal studio-quality dataset with 251 hours from 308 speakers. For English, we use VCTK~\cite{veaux2017superseded}, which contains 44 hours of speech from 110 speakers. The above data are used to train a universal vocoder. The vocoder is trained to synthesize speech from mel-spectrograms generated by different acoustic models, without requiring fine-tuning.

\vspace{-8pt}
\subsubsection{Testing set}
\vspace{-5pt}
We conduct both copy synthesis and Text-to-Speech (TTS) experiments on various types of data.
 
\textbf{Copy synthesis.} We randomly select 10 speakers from the training set as \textit{seen} speakers. Another 10 \textit{unseen} speakers are randomly selected from AISHELL-3~\cite{DBLP:conf/interspeech/ShiBXZL21} and LibriTTS~\cite{DBLP:conf/interspeech/ZenDCZWJCW19}. For each speaker, 50 utterances are adopted to conduct evaluations. Note that ambient noise and slight reverberation are inevitable in AISHELL-3 and LibriTTS due to the recording conditions. 


\textbf{TTS.} For TTS evaluation, we first train a multi-speaker multi-lingual DelightfulTTS~\cite{DBLP:journals/corr/abs-2110-12612} using the same data as the vocoder training, conditioned on speaker ID and the language ID. We evaluate audio quality generated from different TTS tasks to verify the robustness of the universal vocoder. For the seen speakers, we randomly select 10 speakers from the training set. The unseen speakers are the same as the testing set in the copy synthesis test.

We test our approach on the following TTS tasks:
\vspace{-3pt}
\begin{itemize}
\item (a) \textbf{Low-quality speech few-shot}: The few-shot TTS task is evaluated using a low-quality speech dataset, recorded from the mobile phone in a typical office room, with a 2-minute speech per speaker. Two female and three male speakers are involved.
\vspace{-4pt}
\item (b) \textbf{Conversational TTS}: The expressive conversational speech synthesis task is evaluated by a conversation dataset, 10 hours in total, containing 2 speakers.
\vspace{-4pt}
\item (c) \textbf{Unseen styles}: The stylistic speech synthesis task is evaluated by a single-speaker multi-style dataset, 30 minutes in total, with 5 styles, i.e., poetry, fairy tale, joke, story, and thriller.
\vspace{-4pt}
\item (d) \textbf{Emotional TTS}: The emotional speech synthesis task is evaluated by a single-speaker multi-emotion dataset, 12 hours in total, with 6 emotions, i.e., sad, angry, happy, disgusted, fearful, and surprise.
\vspace{-4pt}
\item (e) \textbf{Cross-lingual TTS}: The Cross-lingual TTS task is evaluated by 2 speakers in the training set containing both Chinese and English speech.
\vspace{-4pt}
\item (f) \textbf{Unseen languages}: The performance for languages, unseen to the vocoder training, is evaluated by a Thai dataset containing 2 speakers with 4-hour Thai speech.
\end{itemize}
\vspace{-3pt}
For (a), (b), (c), and (d), we fine-tune the base acoustic model for diverse specific scenarios. For (f), we train another DelightfulTTS model with the base acoustic model's configuration. As the speakers are already included in the base acoustic model, we can directly use English/Chinese text, speaker ID from Chinese/English, and language ID ``English/Chinese" in the acoustic model for (e). 
\vspace{-3pt}

%


\vspace{-10pt}
\subsection{Model Details}
\vspace{-6pt}
We compare our proposed DSPGAN with various popular neural vocoders, including multi-band MelGAN~\cite{DBLP:conf/slt/YangYLF0X21}, HiFi-GAN~\cite{DBLP:conf/nips/KongKB20}, where the former targets a small footprint and low complexity and the latter stands for top-level quality with larger model size.  Accordingly, our DSPGAN has two versions – DSPGAN-mm and DSPGAN-hf, which are based on multi-band MelGAN and HiFi-GAN respectively.  We also compare with NHV~\cite{DBLP:conf/interspeech/LiuCY20}, which is a vocoder based on the source-NNFilter model.



 We use the original model configuration of HiFi-GAN V1, multi-band MelGAN, and NHV in their papers. For our proposed DSPGAN, downsample factors in the downsampling networks are in reverse to the upsample factors in the upsampling networks, and the number of corresponding channels is 32.  We use the pitch extracted from Harvest~\cite{DBLP:conf/interspeech/Morise17} with Stonemask, and the pitch/weight predictor has 4 layers of Conv1D with 256 channels. For the discriminator and training loss of all the above models, we follow the configuration in HiFi-GAN~\cite{DBLP:conf/nips/KongKB20}. Compared with HiFi-GAN/multi-band MelGAN, an additional 1.16/0.82 Gflops computational complexity is assumed in DSPGAN-hf/DSPGAN-mm.  

\vspace{-10pt}
\subsection{Experimental Results}
\vspace{-4pt}
We conduct both objective and subjective evaluations for copy synthesis. For TTS experiments, we synthesize speech from new text and perform subjective evaluation accordingly. We use mel cepstrum distance (MCD) as an objective metric and  mean opinion score (MOS) for subjective listening where 20 listeners are recruited to evaluate the speech quality.

\begin{table*}[tp]
\renewcommand{\thetable}{2}
\vspace{-8pt}
\centering
	\caption{Experimental results in terms of MOS for TTS.}
\resizebox{\linewidth}{!}{ 
\begin{tabular}{l|c|c|c|c|c|c|c|c}
\toprule
      & Seen Spkr& Unseen Spkr & Few-shot & Conversation & Unseen Style & Emotional & Cross-lingual & Unseen Lang \\ \midrule
Multi-band MelGAN  &   3.65$\pm$0.09  & 2.95$\pm$0.08 &2.70$\pm$0.07  & 2.81$\pm$0.10  & 2.58$\pm$0.10 & 2.50$\pm$0.10 &  2.55$\pm$0.12 &  2.78$\pm$0.11    \\
DSPGAN-mm    & 3.98$\pm$0.11 & 3.95$\pm$0.10 & 3.40$\pm$0.08  & 3.89$\pm$0.07 & 3.54$\pm$0.08 & 3.51$\pm$0.09  &   3.71$\pm$0.10 & 3.63$\pm$0.08    \\

HiFi-GAN    & 3.97$\pm$0.10 & 3.25$\pm$0.09 & 2.78$\pm$0.09  & 3.06$\pm$0.08 & 2.79$\pm$0.10 & 2.81$\pm$0.11  &  3.23$\pm$0.10  & 2.94$\pm$0.09    \\
DSPGAN-hf    & \textbf{4.08$\pm$0.07}  & \textbf{3.98$\pm$0.07} &  \textbf{3.58$\pm$0.09} & \textbf{4.04$\pm$0.08}  & \textbf{3.64$\pm$0.08} & \textbf{3.54$\pm$0.10}  & \textbf{3.72$\pm$0.07}   &  \textbf{3.66$\pm$0.08}   \\
NHV    & 2.93$\pm$0.08 & 2.83$\pm$0.08 & 2.75$\pm$0.07  &2.95$\pm$0.09  & 2.64$\pm$0.10 & 2.55$\pm$0.10  &  2.81$\pm$0.11  & 2.90$\pm$0.11   \\
\bottomrule

\end{tabular}
}
\label{MOS test}
\vspace{-12pt}
\end{table*}

\vspace{-10pt}
\subsubsection{Copy synthesis}
\vspace{-5pt}
As shown in Table ~\ref{copysyn test}, multi-band MelGAN and HiFi-GAN face clear quality degradation on re-synthesized speech for unseen speakers. In contrast, degradation from seen to unseen speakers is not obvious for NHV, showing its robustness to unseen speakers. Importantly, DSPGAN-mm/DSPGAN-hf performs much better on unseen speakers compared with their counterparts, thanks to the robustness that benefited from the source-NNFilter model's supervision. 

\vspace{-12pt}
\begin{table}[ht]
\renewcommand{\thetable}{1}
\centering
    \vspace{-2pt}
	\caption{Results on copy synthesis in terms of MOS \& MCD.}
\resizebox{0.95\linewidth}{!}{ 
\begin{tabular}{l|cc|cc}
\toprule
      & \multicolumn{2}{c|}{Seen speakers} & \multicolumn{2}{c}{Unseen speakers} \\ \midrule
Model & MOS    & MCD    & MOS     & MCD   \\ \midrule
Multi-band MelGAN     & 3.89$\pm$0.07 & 3.97 &3.45$\pm$0.06 & 4.38   \\
DSPGAN-mm & 4.01$\pm$0.07 &3.03 & 3.96$\pm$0.05 & 3.11    \\
HiFi-GAN    &  4.23$\pm$0.05 & 2.33 & 3.84$\pm$0.06 & 2.59   \\
DSPGAN-hf     & \textbf{4.24$\pm$0.06} & \textbf{2.31} & \textbf{4.10$\pm$0.06} &  \textbf{2.36}   \\
NHV     & 3.88$\pm$0.07 & 2.59 & 3.81$\pm$0.06 & 2.52    \\
\midrule
Recording     & 4.51$\pm$0.06 & - & 4.48$\pm$0.06 & -   \\ \bottomrule

\end{tabular}
}
\vspace{-10pt}
\label{copysyn test}
\renewcommand{\thefigure}{1}
\end{table}

\vspace{-12pt}
\subsubsection{TTS}
\begin{figure}[ht]
\setlength{\abovecaptionskip}{0.1cm}
	\centering
	\includegraphics[width=1\linewidth]{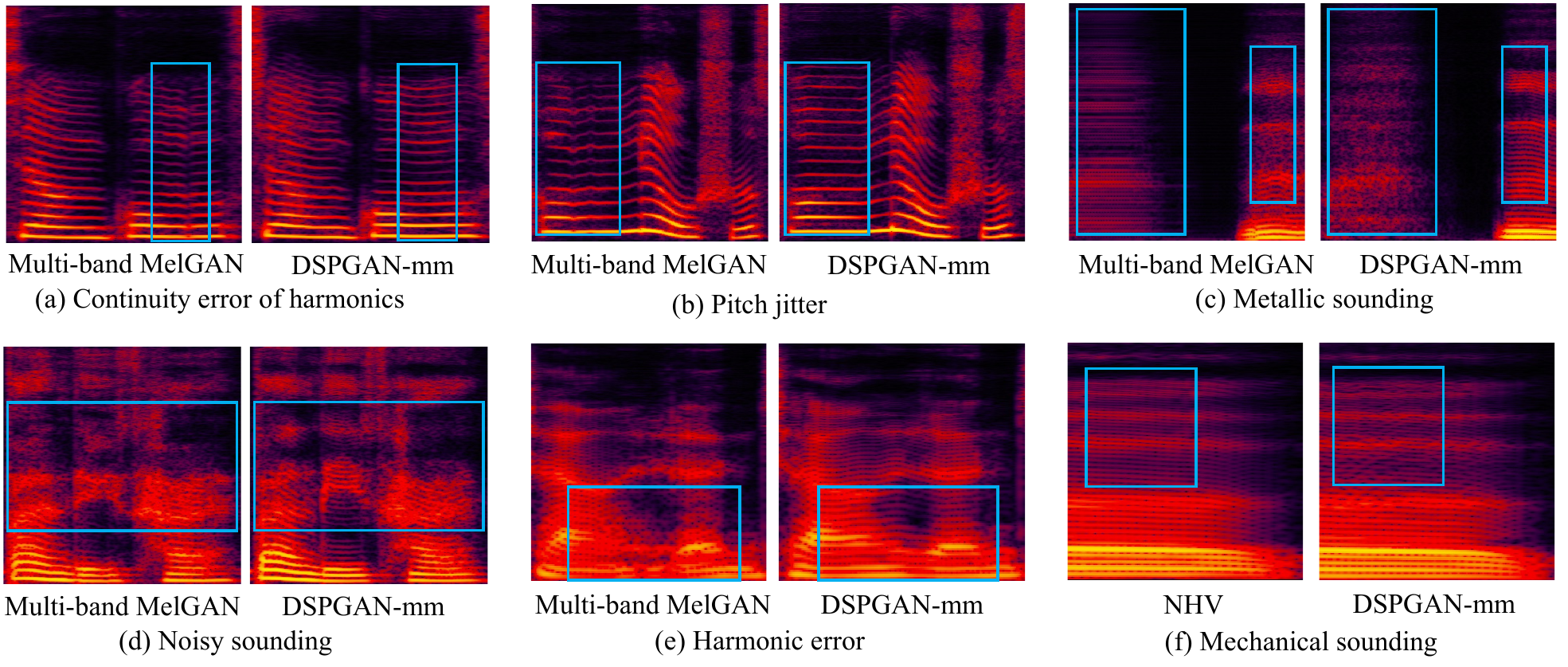}
	\vspace{-16pt}
	\caption{
		The spectrograms of synthesized samples from DSPGAN-mm, multi-band MelGAN and NHV.
	}
	\label{artifect}\vspace{-14pt}
	
\end{figure}

\vspace{-6pt}
For the TTS experiments shown in Table~\ref{MOS test}, DSPGAN-mm and DSPGAN-hf show their superiority compared to other systems in different testing scenarios. An interesting finding is the overall poor performance achieved by the model of NHV directly used in waveform generation, which indicates the effectiveness of the supervision by NHV for the proposed approaches, rather than the good performance for the audio reconstruction of NHV.
To intuitively present the advantages of the proposed DSPGAN for avoiding artifacts, spectrograms from DSPGAN-mm, multi-band MelGAN and NHV are displayed in Figure~\ref{artifect}. As shown in (a), (b), and (e), DSPGAN can avoid discontinuous harmonics, pitch jitter, and harmonic errors, due to the introduction of sine excitation. And in (c) and (d), since extracted mel-spectrogram from the source-NNFilter model is used as time-frequency domain supervision in DSPGAN, metallic and noisy soundings are clearly avoided. In addition, as can be seen in (f), compared with NHV, DSPGAN has more aperiodic components at high frequency instead of only harmonics, which benefits from the good modeling ability of the GAN-based vocoder and the time-frequency domain supervision. 



\vspace{-10pt}

\vspace{-8pt}
\subsection{Ablation Study}
\vspace{-5pt}
We conduct an ablation study to investigate the effectiveness of designs in DSPGAN, including time domain supervision, time-frequency domain supervision, and weight predictor. As shown in Table~\ref{MOS for ab}, the MOS scores demonstrate that individually removing the three components would lead to audio quality degradation. These results indicate the advantages of the proposed approach in universal audio reconstruction.

\vspace{-10pt}
\begin{table}[ht]
    
	\centering
	\vspace{-5pt}
	\caption{The MOS test of ablation study. T-/TF-S denotes time/time-frequency domain supervision.}
	\resizebox{0.55\linewidth}{!}{
	\begin{tabular}{lcl}
		\toprule
		Model &MOS  \\ 
		\midrule
		
		DSPGAN-mm &3.93$\pm$0.05  \\
        \ w/o T-S    &3.62$\pm$0.06  \\
        \ w/o TF-S    &3.08$\pm$0.05 \\
        \ w/o weight predictor &3.77$\pm$0.05 \\
        
		\bottomrule
	\end{tabular}
	}
	\label{MOS for ab}
	\vspace{-8pt}
\end{table}

\vspace{-12pt}
\section{Conclusions}
\label{sec:Conclusions}
\vspace{-7pt}
In this paper, we propose DSPGAN, aiming to build a universal vocoder by introducing the time-frequency domain supervision from DSP for the GAN-based vocoder. With the supervision of mel-spectrogram extracted from the signal of the DSP module and sine excitation, DSPGAN effectively eliminates the mechanical sounding of the DSP vocoder and artifacts of the GAN-based vocoder. Experimental results show that the proposed universal DSPGAN achieves good performance on various acoustic models, speakers, languages, and speaking styles to synthesize high-fidelity waveforms.

\vfill\pagebreak

\bibliographystyle{IEEE}
\bibliography{strings,refs}

\end{document}